\documentclass[fleqn,usenatbib]{mnras}
\usepackage[T1]{fontenc}
\DeclareRobustCommand{\VAN}[3]{#2}
\let\VANthebibliography\thebibliography
\def\thebibliography{\DeclareRobustCommand{\VAN}[3]{##3}\VANthebibliography}
\usepackage{graphicx}	
\usepackage{amsmath}	
\usepackage{amssymb}

\usepackage{newtxtext,newtxmath}
\usepackage{lscape}
\usepackage{ulem}

\title[$\omega$\,Cen: outer fields catalogues]{
The \textit{HST} large programme on $\omega$\,Centauri -- IV. catalogue of two external fields}

\author[Scalco et al.]{M.\,Scalco$^{1,2}$\thanks{E-mail: michele.scalco@unife.it},
A.\,Bellini$^3$, 
L.\,R.\,Bedin$^2$, 
J.\,Anderson$^3$, 
P.\,Rosati$^1$,  
M.\ Libralato$^4$,
\newauthor 
M.\,Salaris$^5$,  
E.\,Vesperini$^6$,
D.\,Nardiello$^{7,3}$,
D. Apai$^{8,9}$,
A. J. Burgasser$^{10}$
\newauthor 
and R. Gerasimov$^{10}$.
\\
$^{1}$Dipartimento di Fisica e Scienze della Terra, Università di Ferrara, Via Giuseppe Saragat 1, Ferrara I-44122, Italy\\
$^{2}$Istituto Nazionale di Astrofisica, Osservatorio Astronomico di Padova, Vicolo dell’Osservatorio 5, Padova I-35122, Italy\\
$^{3}$Space Telescope Science Institute, 3700 San Martin Drive, Baltimore, MD 21218, USA\\ 
$^{4}$AURA for the European Space Agency (ESA), ESA Office, Space Telescope Science Institute, 3700 San Martin Drive, Baltimore, MD 21218, USA\\
$^{5}$Astrophysics Research Institute, Liverpool John Moores University, IC2, Liverpool Science Park, 146 Brownlow Hill, Liverpool, L3 5RF, UK\\
$^{6}$Department of Astronomy, Indiana University, Bloomington, Swain West, 727 E. 3rd Street, IN 47405, USA\\ 
$^{7}$Aix Marseille Univ., CNRS, CNES, LAM, Marseille, France\\
$^{8}$Department of Astronomy and Steward Observatory, The University of Arizona, 933 N. Cherry Avenue, Tucson, AZ 85721, USA\\
$^{9}$Lunar and Planetary Laboratory, The University of Arizona, 1629 E. University Blvd., Tucson, AZ 85721, USA\\
$^{10}$Center for Astrophysics and Space Science, University of California San Diego, La Jolla, CA 92093, USA\\
}

\date{Accepted 2021 May 18. Received 2021 May 18; in original form 2021 April 20}

\pubyear{2021}

\begin{document}
\label{firstpage}
\pagerange{\pageref{firstpage}--\pageref{lastpage}}
\maketitle

\begin{abstract}
In the fourth paper of this series, we present --and publicly release-- the state-of-the-art catalogue and atlases for the two remaining parallel fields observed with the \textit{Hubble Space Telescope} for the large programme on $\omega$\,Centauri. These two fields are located at $\sim$12$^\prime$ from the centre of the globular cluster (in the West and South-West directions) and were imaged in filters from the ultraviolet to the infrared. Both fields were observed at two epochs separated by about 2\,years that were used to derive proper motions and to compute membership probabilities.  
\end{abstract}

\begin{keywords}
star clusters: individual : $\omega$\,Centauri (NGC5139) - stars: Hertzsprung–Russell and colour–magnitude diagrams - Population II - techniques: photometric - catalogues
\end{keywords}


\section{Introduction}
The “\textit{Hubble Space Telescope} (\textit{HST}) large program of $\omega$ Centauri” (GO-14118 + GO-14662, PI: Bedin, L. R.) aims at observing the white dwarf (WD) cooling sequence (CS) for the stars of this Galactic globular cluster (GC) down to the faintest WDs. These observations aim to better characterize the multiple WD CSs discovered within this cluster (Bellini et al.\, 2013) and to investigate the connection between the WD CSs with the well-known main sequence (MS) multiple populations (mPOPs) (Bedin et al.\ 2004, Villanova et al.\ 2007, Bellini et al.\ 2009, 2010, 2017c, 2018, Marino et al.\ 2011, Milone et al.\ 2017) and the cluster Helium enhancement (Norris 2004, King et al.\ 2012).

The primary data-set of the program includes observations of a primary field (hereafter, field F0) obtained with the Wide-Field Channel (WFC) of the Advanced Camera for Surveys (ACS), located about 12$^\prime$ from the cluster’s centre. F0 is the only field that is sufficiently deep to reach the faintest theoretically detectable WDs in the cluster. In order to de-contaminate the fields from background and foreground objects, the pair of programs (GO-14118 and GO-14662) was designed to obtain observations at two epochs. For each epoch, the main field F0 was observed in 66 orbits and at 3 different orientations (22 orbits each), with the aim of minimizing the impact of imperfect Charge Transfer Efficiency (CTE) correction, imperfect calibrations, and systematic errors. 

Given that specific orientations are available only at different epochs of the year, \textit{de facto}, each of the three orientations is a sub-epoch for the main F0 field, but it also places the parallel observations in three additional and different fields. These parallel fields were taken with the Wide-Field-Camera\,3 (WFC3) in both the Ultraviolet-Visible (UVIS) channel (8 orbits per epoch per field) and in the Infrared (IR) channel (14 orbits per epoch per field).

These three parallel fields (hereafter referred to as F1, F2, and F3) were collected with the purpose of studying, through different approaches, the mPOPs in stars at different evolutionary phases and at different radial distances from the cluster centre. 
Parallel fields were observed with different filters in order to have a wider colour baseline to identify and better separate the different mPOPs within $\omega$\,Centauri (hereafter, $\omega$\,Cen).

The exposures from the parallel field F1 were reduced and presented in the three previous publications of this series: the mPOPs at very faint magnitudes were analysed by Milone et al.\ (2017, Paper\,I). Bellini et al.\ (2018, Paper\,II) analysed the internal kinematics of the mPOPs, complementing our GO14118 + GO-14662 data with archival images collected more than 10\,yrs earlier under \textit{HST} programs GO-9444 and GO-10101 (on both PI: King). Finally, Libralato et al.\ (2018, Paper\,III) presented the absolute proper motion estimate for $\omega$ Cen in our field F1.

In this paper, we present and release the catalogue and the atlases for the two remaining WFC3 parallel fields, F2 and F3. Our new catalogue provides multi-band photometry, proper motions (PMs), and membership probabilities for all sources detected in our fields; the atlases are high-resolution FITS images, with headers containing the astrometric solutions with keywords in the World Coordinate System (WCS).

This paper is organised as follows: 
Section 2 is dedicated to the description of the data; 
Section 3 briefly outlines the data reduction process; 
Section 4 presents some of the colour-magnitude diagrams obtained; 
Section 5 describes the PM measurements and the methodology to estimate membership probability; 
in Section 6, we describe in details the content of the data released tables. 
Finally, in Section 7 we briefly summarize the key results, indicating potential immediate future uses of this catalogue and also identify our own upcoming scientific investigations that will make use of it.

\section{Data set}
\begin{centering} 
 \begin{figure}
\includegraphics[width=8cm]{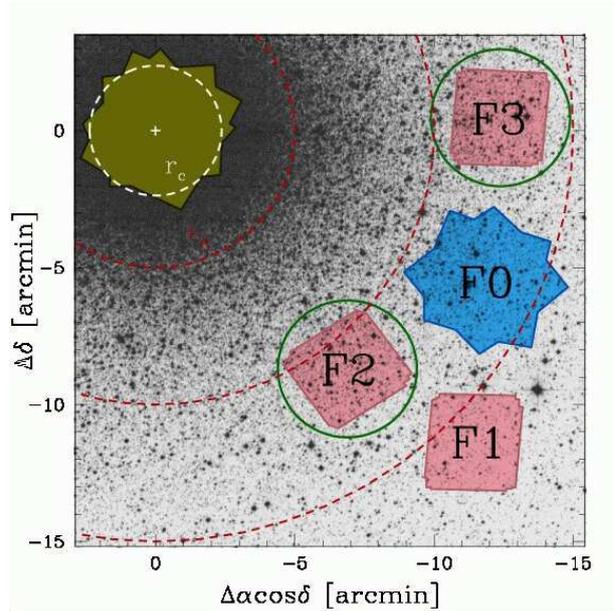}
  \caption{Outlines of the fields observed in \textit{HST} programs GO-14118 + GO-14662, superimposed on a DSS image of $\omega$ Cen. The primary ACS/WFC field (F0) is in azure, while the three parallel WFC3 fields are shown in pink. We also show, in yellow, the central field presented in Bellini et al.\ (2017a, b, c). Units are in arcmin measured from the cluster centre. The data discussed in this paper come from fields F2 and F3, which are marked with green circles. The white and red dashed circles mark the cluster's core radius ($r_{\rm c}=2^\prime_{\cdot}37$), the half-light radius ($r_{\rm h}=5^\prime_{\cdot}00$), at $2\,r_{\rm h}$ and $3\,r_{\rm h}$, respectively, from the centre.}
  \label{fig:fields} 
 \end{figure} 
 \end{centering}

Fields F2 and F3 were observed in 2016 (GO-14118) and 2018 (GO-14662), using both channels of the WFC3. In each epoch, data were collected with the UVIS channel in five filters (F275W, F336W, F438W, F606W, and F814W) and with the IR channel in two filters (F110W and F160W). Table 1 reports the complete list of \textit{HST} observations of fields F2 and F3 for each epoch. \\
Figure \ref{fig:fields} shows the locations of the GO-14118 and GO-14662 fields (F0 to F3), superimposed on an image from the Digital Sky Survey (DSS)\footnote{https://archive.eso.org/dss/dss}. The primary ACS/WFC field (F0) is shown in azure, while the three parallel WFC3 fields (F1 to F3) are plotted in pink. For reference, we also show the central field (in yellow) analysed in Bellini et al.\ (2017a, b, c). The GO-14118 and GO-14662 fields cover a radial extent from $\sim 2\,r_{\rm h}$ to $\sim 4\,r_{\rm h}$ (being $r_{\rm h}=5^\prime_{\cdot}00$ the half-light radius, Harris 1996). In this article, we consider only data for fields F2 and F3 (both circled in green).

\begin{table*}
\textbf{Table 1}\\
List of \textit{HST} Observations of Fields F2 and F3\\
\smallskip
\begin{tabular}{c c c} 
 & Field F2 & \\
 \hline
 \hline
 Filter & Exposures & Epoch\\
 \hline
 \hline
 & Epoch 1 (GO-14118) &\\
 \hline
  & WFC3/UVIS &\\
 \hline
 F275W & 4$\times$1328 s & 2016/07/01-05\\ 
 F336W & 4$\times$1230 s & 2016/07/01-05\\ 
 F438W & 4$\times$98 s & 2016/07/01-05\\
 F606W & 2$\times$99 s + 2$\times$1255 s + 2$\times$1347 s & 2016/06/30-07/04\\
 F814W & 2$\times$98 s + 2$\times$1253 s + 2$\times$1345 s & 2016/06/27-07/04\\
 \hline
  & WFC3/IR &\\
 \hline
 F110W & 7$\times$143 s + 14$\times$1303 s & 2016/06/24-07/04\\
 F160W & 7$\times$143 s + 14$\times$1303 s & 2016/06/27-07/05\\
 \hline
 & Epoch 2 (GO-14662) &\\
 \hline
   & WFC3/UVIS & \\
 \hline
 F275W & 1$\times$1240 s + 3$\times$1243 s & 2018/06/30-/07/01\\
 F336W & 4$\times$1157 s & 2018/06/30-/07/01\\
 F438W & 4$\times$104 s & 2018/06/30-/07/01\\
 F606W & 2$\times$104 s + 2$\times$1186 s + 2$\times$1266 s & 2018/06/30\\
 F814W & 2$\times$104 s + 2$\times$1186 s + 2$\times$1266 s & 2018/06/28\\
 \hline
  & WFC3/IR &\\
 \hline
 F110W & 7$\times$143 s + 14$\times$1203 s & 2018/06/24-07/01\\
 F160W & 7$\times$143 s + 14$\times$1203 s & 2018/06/27-30\\
 \hline
 \end{tabular}
 \begin{tabular}{c c c} 
  & Field F3 & \\
 \hline
 \hline
 Filter & Exposures & Epoch\\
 \hline
 \hline
 & Epoch 1 (GO-14118) &\\
 \hline
  & WFC3/UVIS & \\
 \hline
 F275W & 4$\times$1328 s & 2016/01/31\\ 
 F336W & 4$\times$1230 s & 2016/01/31\\ 
 F438W & 4$\times$98 s & 2016/01/31\\
 F606W & 2$\times$99 s + 2$\times$1255 s + 2$\times$1347 s & 2016/01/30\\
 F814W & 2$\times$98 s + 2$\times$1253 s + 2$\times$1345 s & 2016/01/30-31\\
 \hline
  & WFC3/IR & \\
 \hline
 F110W & 7$\times$143 s + 14$\times$1303 s & 2016/01/30-02/04\\
 F160W & 7$\times$143 s + 14$\times$1303 s & 2016/02/04-05\\
 \hline
 & Epoch 2 (GO-14662) &\\
 \hline
   & WFC3/UVIS & \\
 \hline
 F275W & 4$\times$1229 s & 2018/01/30-31\\
 F336W & 4$\times$1143 s & 2018/01/30-31\\
 F438W & 4$\times$104 s & 2018/01/30-31\\
 F606W & 1$\times$95 + 1$\times$104 s + 2$\times$1172 s + 2$\times$1252 s & 2018/01/30\\
 F814W & 2$\times$104 s + 2$\times$1172 s + 2$\times$1252 s & 2018/01/30-31\\
 \hline
  & WFC3/IR & \\
 \hline
 F110W & 7$\times$143 s + 14$\times$1203 s & 2018/02/09-24\\
 F160W & 7$\times$143 s + 14$\times$1203 s & 2018/02/20-25\\
 \hline
\end{tabular}
\end{table*}

\section{Data Reduction}
In this analysis, we used only \texttt{\_flc}-type images (in units of e$^-$) for WFC3/UVIS, and \texttt{\_flt}-type images (in units of e$^-$/s) for WFC3/IR, as they preserve the pixel data with its original sampling for stellar-profile fitting.  

These \texttt{\_flc}-type and \texttt{\_flt}-type images are both corrected via standard calibrations (bias, flat field and dark); in addition, the \texttt{\_flc}-type images are also corrected for CTE defects following the empirical approach described in Anderson \& Bedin (2010). 

\subsection{First-Pass Photometry}
We measure the stellar positions and fluxes in each exposure using the \texttt{FORTRAN} code \texttt{hst1pass}, which is a generalised version of the \texttt{img2xym\_WFC} software package (Anderson \& King 2006). Starting from spatially-variable --but time-averaged-- empirical Point Spread Function (PSF) libraries (e.g., Anderson \& King 2006), the routine \texttt{hst1pass} runs a single pass of source finding for each exposure without performing neighbor subtraction. We perturbed the empirical PSF in order to find the best spatially variable PSF for each image. Stellar positions and fluxes are obtained by fitting each source with the obtained PSF. Stellar positions in each single-exposure catalogue are corrected for geometric distortion by using the state-of-the-art geometric-distortion corrections of Bellini et al.\, (2011) for WCF3/UVIS, and the publicly available WFC3/IR correction developed by J.\, Anderson (Anderson 2016, Instrument Science Report WFC3 2016-12, Appendix\,A) \footnote{Available at https://www.stsci.edu/~jayander/STDGDCs/}. 

\subsection{The Master Frame}
For each field, we defined a common, pixel-based reference coordinate system, based on a WFC3/UVIS F814W single-exposure catalogue. Then, for the images taken in each filter, we used only bright, unsaturated, and well-measured stars to derive general six-parameter linear transformations to transform stellar positions -- as measured in each individual exposure -- onto the common reference frame system. The photometry of these preliminary catalogues was zero-pointed to the first long exposure taken in each filter/epoch. 

\subsection{Second-Pass Photometry}
The second-pass photometry is performed through the \texttt{FORTRAN} software package \texttt{KS2}, which is based on \texttt{kitchen\_sync}, originally designed to reduce specific ACS/WFC data (Anderson et al.\ 2008). The code \texttt{KS2}, also developed by J.\, Anderson, takes images, perturbed PSF arrays, and transformations obtained during the “first-pass photometry” stage to simultaneously find and measure stars in all of the individual exposures and for the entire set of filters. By relying on multiple exposures, \texttt{KS2} finds and measures faint stars that would be otherwise lost in the noise of individual exposures. A detailed description of Anderson's code is given in Bellini et al.\, (2017a) and Nardiello et al.\ (2018).

The star-finding procedure is accomplished through different passes of finding, moving progressively from the brightest to the faintest stars. During the initial star-finding pass, the software starts from a list of bright stars, available from the first-pass photometry, and constructs weighted masks around the bright stars, which helps the software avoid PSF-related artifacts. Then, \texttt{KS2} subtracts the bright stars. In the following pass, the routine searches for stars that are fainter than the stars from previous iteration, and then measures and subtracts them. In each successive iteration of finding, \texttt{KS2} identifies stars that satisfy increasingly relaxed search criteria. 

For this project, we chose to execute nine iterations of finding. To make the catalogue as similar as possible to that of the F1 field released by Bellini et al.\  (2018), we performed the star-finding using the F606W and F814W filters. In the first four iterations we required that a star be present in the F606W and F814W long exposures. In the last pass we focused on the short exposures to get \texttt{KS2}-derived photometry for the brighter stars.

\texttt{KS2} has three approaches for measuring stars, each of which is best-suited for stars in different magnitude regimes. The first method gives the best results for stars that are bright enough to generate a high signal-to-noise peak within its local 5$\times$5 pixel, neighbour-subtracted raster. When that happens, the routine measures, in each image, the flux and the position of the source using an appropriate local PSF, after subtracting the neighbour stars. The local sky value is computed using the surrounding pixels in an annulus (between 5 and 8 pixels in radius), with the contributions of the neighbors and the star itself subtracted.

Methods two and three work best for faint stars and in crowded environments. In method two, starting from the position obtained during the finding stage, \texttt{KS2} uses the PSF to determine a best-fit flux from the inner 3$\times$3 pixels. Method three is similar, but it uses the brightest 4 pixels and weights them by the expected fraction of the PSF in those pixels. For a detailed description of the three methods, we refer to Bellini et al.\ (2017a) and Nardiello et al.\ (2018). We have verified that the photometry of the three methods for stars near the overlap magnitude regions is consistent. 

Saturated stars are not measured by \texttt{KS2}. However, their position and fluxes are recovered from the first-pass photometry and supplemented in output. Our final photometric catalogue contains a total of 42,551 sources in both fields.

In addition to the astro-photometric catalogue, \texttt{KS2} outputs stacked 
images obtained from the \texttt{\_flc} and \texttt{\_flt} exposures. For each field, we generated 11 different stacked images: one for the filters F275W, F336W, and F438W, and two for the filters F606W, F814W, F110W, and F160W, separating short- and long-exposure images. We make these stacked image pairs available with 1$\times$1 and 2$\times$2 pixel super-sampling. 

\begin{centering} 
 \begin{figure*}
  \includegraphics[width=15cm]{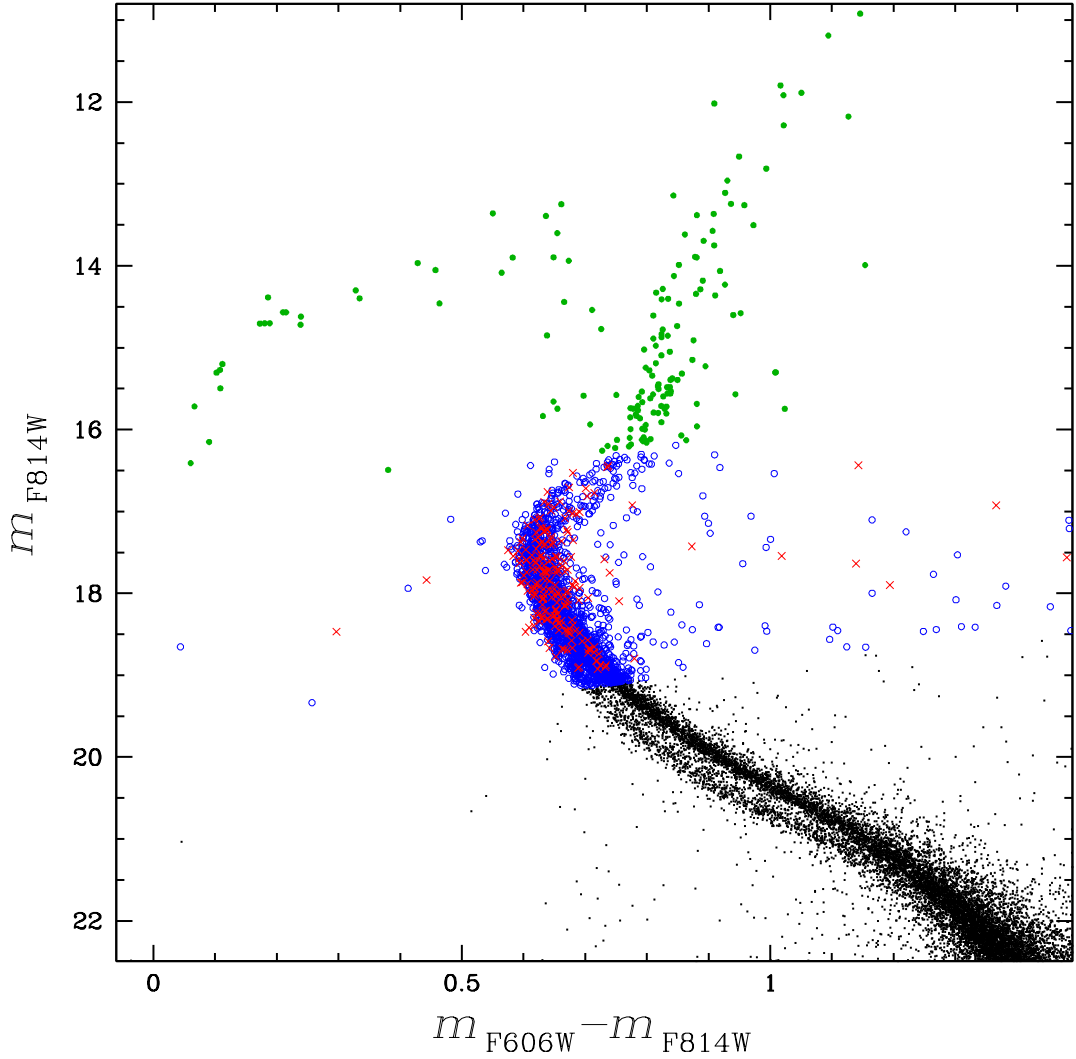}
  \caption{Bright part of the $m_{\rm F814W}$ vs. $m_{\rm F606W}-m_{\rm F814W}$ CMD. The unsaturated stars in the long exposures for both filters (black dots) and stars saturated in the long exposure but not in the short ones (blue circles) are directly measured by \texttt{KS2}. Red crosses represent stars that are saturated in at least one filter long exposure and not measured in the short ones. Finally, green dots represent stars that are saturated in at least one filter short exposure. The fluxes of those stars are measured in the first-pass photometry.}
  \label{fig:saturated} 
 \end{figure*} 
 \end{centering}

Figure \ref{fig:saturated} shows the upper part of the $m_{\rm F814W}$ vs. $m_{\rm F606W}-m_{\rm F814W}$ colour-magnitude diagram (CMD). Black dots are unsaturated stars in both the F606W and F814W long exposures, while red crosses mark stars that are saturated in the long exposure in at least one filter and were not found in the short exposures. Finally, stars marked with blue circles are saturated in the long exposures but are found unsaturated in the short ones. Stars marked with green dots are stars saturated in at least one filter in the short exposures. Black dots and blue circles are stars deemed best-measured by method one in \texttt{KS2}, while the positions and the fluxes of the stars marked with red crosses and green dots are available through the first-pass photometry, since saturated stars are not measured by \texttt{KS2}.

\subsection{Photometric calibration}
The photometry has been zero-pointed into the Vega magnitude system by following the recipe of Bedin et al.\ (2005), Bellini et al.\ (2017a) and Nardiello et al.\ (2018). The process of zero-pointing \textit{HST}'s photometry is based on the comparison between our PSF-based instrumental magnitudes and the aperture-photometry on \texttt{\_drc} exposures (calibrated and resampled images normalized to 1\,s exposure time). The calibrated magnitude $m_{\text{CAL,X}}$ of a star in the filter X is given by:
\begin{center}
$m_{\text{CAL,X}}$ = $m^{\text{flc}}_{\text{PSF,X}}$ + ZP$_{\text{X}}$ + $<\delta m>$
\end{center}
where $m^{\text{flc}}_{\text{PSF,X}}$ is our instrumental PSF-based magnitude as measured on \texttt{\_flc} (or \texttt{\_flt} for the IR channel) exposures, ZP$_{\text{X}}$ is the filter Vega-mag zero point and $<\delta m>$ is the median magnitude difference between $m^{\text{drc}}_{\text{AP(r,$\infty$)}}$, the aperture photometry measured on \texttt{\_drc} (or \texttt{\_drz}) exposures within a finite-aperture radius $r$ but corrected to account for an infinite-aperture radius and our PSF-based instrumental magnitudes. ZP$_{\text{X}}$ and the encircled energy fractions as a function of $r$ can be found on the WFC3 webpage\footnote{https://www.stsci.edu/hst/instrumentation/wfc3/data-analysis/photometric-calibration} for tabulated wavelengths. 

Since the two fields are crowded, we used only bright, unsaturated stars for the photometric calibration. For this reason we made use of the photometry obtained with the method one, which is best-suited for bright stars.

We measured the \texttt{\_drc} (\texttt{\_drz}) aperture photometry of bright, relatively isolated, and unsaturated stars by using an aperture value of 10 and 3 pixels (0.4 arcsec) for UVIS and IR, respectively. Each of these measurements was then corrected for the finite aperture. For each filter, we cross-identified stars in common between the \texttt{\_drc}-based aperture photometry and our \texttt{KS2} method-one photometry. For each measurement, we then computed the 2.5$\sigma$-clipped median values $<\delta m>$ = $m^{\text{drc}}_{\text{AP(r,$\infty$)}}$ - $m^{\text{flc}}_{\text{PSF,X}}$.  

Finally, we verified that the photometric zero-points evaluated by using methods 2 and 3 are consistent with the values obtained with method 1. Therefore, we apply the calibration correction obtained for method one to the other two methods.  Table 2 summarizes the aperture-correction $<\delta m>$ values obtained for each filter, together with the respective Vega-mag zero points from the STScI website. 

\begin{table}
\begin{centering}
\textbf{Table 2}\\
Photometric-calibration Zero Points\\
\smallskip
\end{centering}
\begin{tabular*}{.48\textwidth}{@{\extracolsep{\fill}}l c c}
\hline      
\hline
Filter & $<\delta m>$ & ZP(Vegamag)\\
& (mag) & (mag)\\
\hline
 & WFC3/UVIS &\\
\hline
F275W & $+7.5965\pm0.04$ & $+22.737$\\
F336W & $+7.5371\pm0.04$ & $+23.554$\\
F438W & $+4.8644\pm0.03$ & $+24.999$\\
F606W & $+7.6494\pm0.07$ & $+25.995$\\
F814W & $+7.6288\pm0.06$ & $+24.684$\\
\hline
 & WFC3/IR &\\
\hline
F110W & $-0.0833\pm0.01$ & $+26.042$\\
F160W & $-0.0916\pm0.02$ & $+24.662$\\
\hline
\end{tabular*}
\end{table}

\subsection{Astrometry}
We cross-reference the stars in our catalogue with the stars in the Gaia early Data Release 3 (Gaia eDR3; Gaia Collaboration et al.\ 2020). Gaia's positions were evolved to the observed epochs. We found about 3,200 sources in common, which were used to anchor our positions (X,Y) to the Gaia eDR3 absolute astrometric system. As such, the positions are referred to the reference epoch of Gaia catalogue, 2016.0, which are in the International Celestial Reference System (ICRS). 

\subsection{Quality parameters}

In addition to positions and fluxes, \texttt{KS2} provides other diagnostic parameters, such as the RMS of the individual-exposure photometry measurements. This latter is useful when selecting the best-measured stars in investigations that require high-precision  stellar evolutionary sequences in CMDs. 

The quality-of-fit (QFIT) parameter informs about the accuracy of the PSF-fitting during the measurements of the position and the flux of a star. The closer to unity the QFIT is, the more a source resembles the adopted PSF model. This parameter allows us to distinguish between stars that are isolated and/or well-measured, and other sources for which the light profiles are not accurately fit by the PSF (cosmic rays, hot pixels, extended sources, blends, etc). 

The “o” parameter is defined as the initial (i.e., before neighbour subtraction) ratio of the light within the fitting radius due to nearby neighbors to the light of the star. Since neighbor subtraction is never perfect, and it is hard to measure faint stars surrounded by much brighter sources, the photometry of sources with large values of the "o" parameter is likely less accurate. 

The parameter RADXS (Bedin et al.\ 2008) is a shape parameter that measures how much the deviation of the PSF shape is from the predictions by comparing the source flux just outside the PSF core and the flux expected from the PSF-model. Galaxies have large positive values of RADXS, while objects sharper than the PSF, e.g. cosmic rays or hot pixels, have large negative values of RADXS.

Finally, \texttt{KS2} also reports the number of images in which a star was found (N$_{\text{f}}$), and the number of good measurements  of the star used to compute its average position and flux (N$_{\text{g}}$)
(those consistent with the average, see Anderson et al.\, 2008 for further details).

\section{colour-magnitude diagrams}

\begin{centering} 
 \begin{figure*}
  \includegraphics[width=15cm]{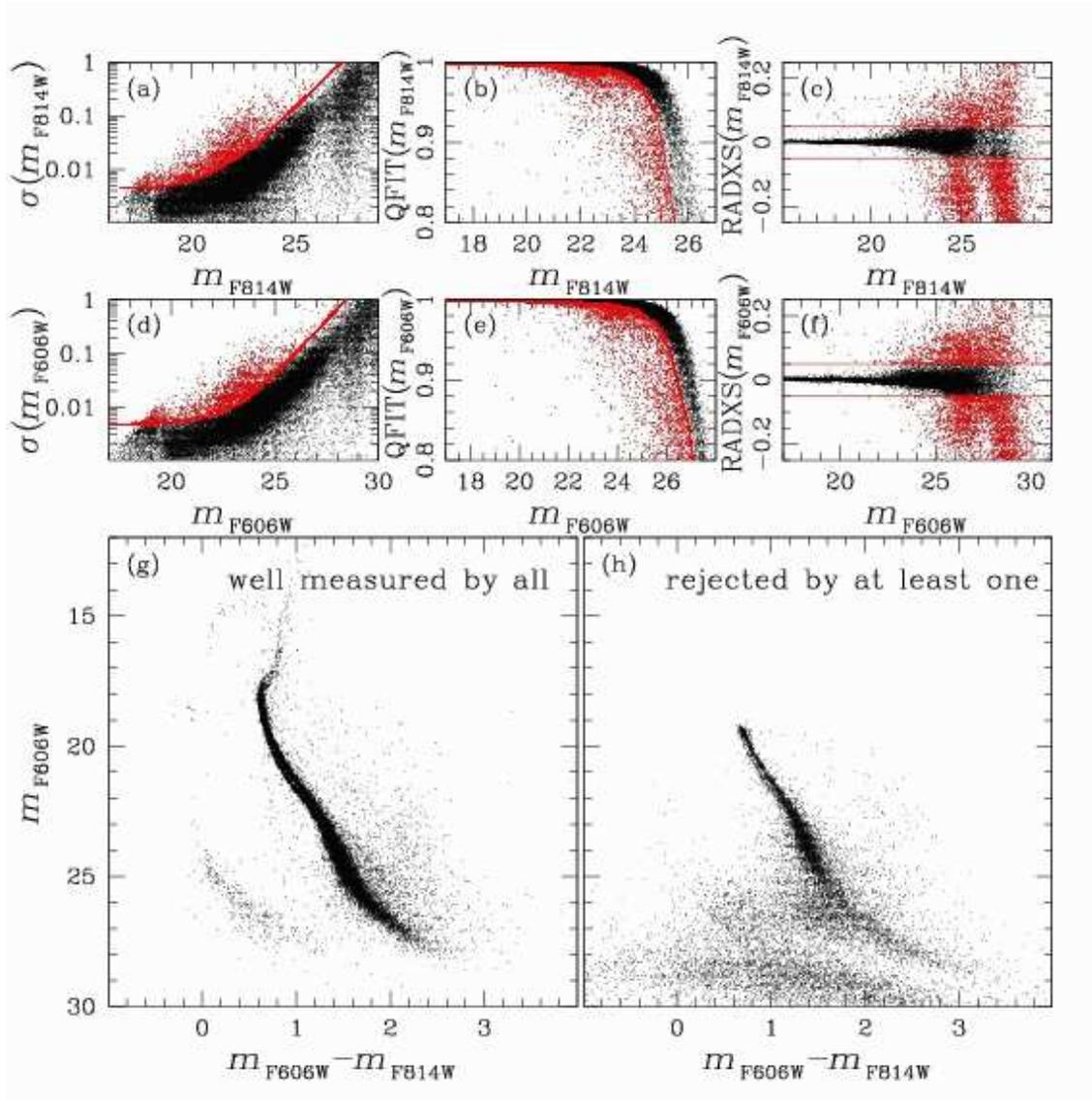}
  \caption{Effect of simple stellar selections based on $\sigma$, QFIT and RADXS. (a)-(c) Selection of the stars based on $\sigma$, QFIT and RADXS in function of the F814W magnitude. The red lines separate the bulk of those defined as well-measured stars from the outliers. The rejected stars are represented in red. (d)-(f) Analogues to (a)-(c) but for F606W photometry. (g) $m_{\rm F606W}$ vs. $m_{\rm F606W}-m_{\rm F814W}$ CMD of stars that are well measured according to all the 6 parameters. (h) Same CMD of (g) but of stars that are rejected by at least one filter.}
  \label{fig:quality_sel} 
 \end{figure*} 
 \end{centering}

In Figure \ref{fig:quality_sel}, we show an example of selection of well-measured stars using the quality parameters provided by \texttt{KS2}. Top and middle panels of Figure \ref{fig:quality_sel} show, respectively from the left to the right, the photometric errors $\sigma$, defined as the RMS divided by the square root of N$_g$, QFIT and RADXS as a function of F814W and F606W magnitudes obtained with method 1 (see Sect. 3). Similar plots can be made using method-2 and method-3 outputs. In this example, the selection criteria for parameters $\sigma$ and QFIT are made by eye, arbitrarily defining a line (indicated in red) that separate the bulk of well-measured stars from the outliers. For the RADXS parameter, we selected all stars that satisfy the condition: $-$0.05 $<$RADXS$<$ $+$0.05 (panels (c) and (f) of Figure \ref{fig:quality_sel}).

The bottom panels of Figure \ref{fig:quality_sel} show the $m_{\rm F606W}$ vs. $m_{\rm F606W}-m_{\rm F814W}$ colour-magnitude diagram for the stars that pass the selection criteria in both filters (panel (g)) and for the stars that were rejected in at least one filter (panel (h)). From the CMDs, it is clear that many stars ($\sim 39\%$) with poor photometric quality are rejected with these tight selections.

\begin{centering} 
 \begin{figure*}
  \includegraphics[width=15cm]{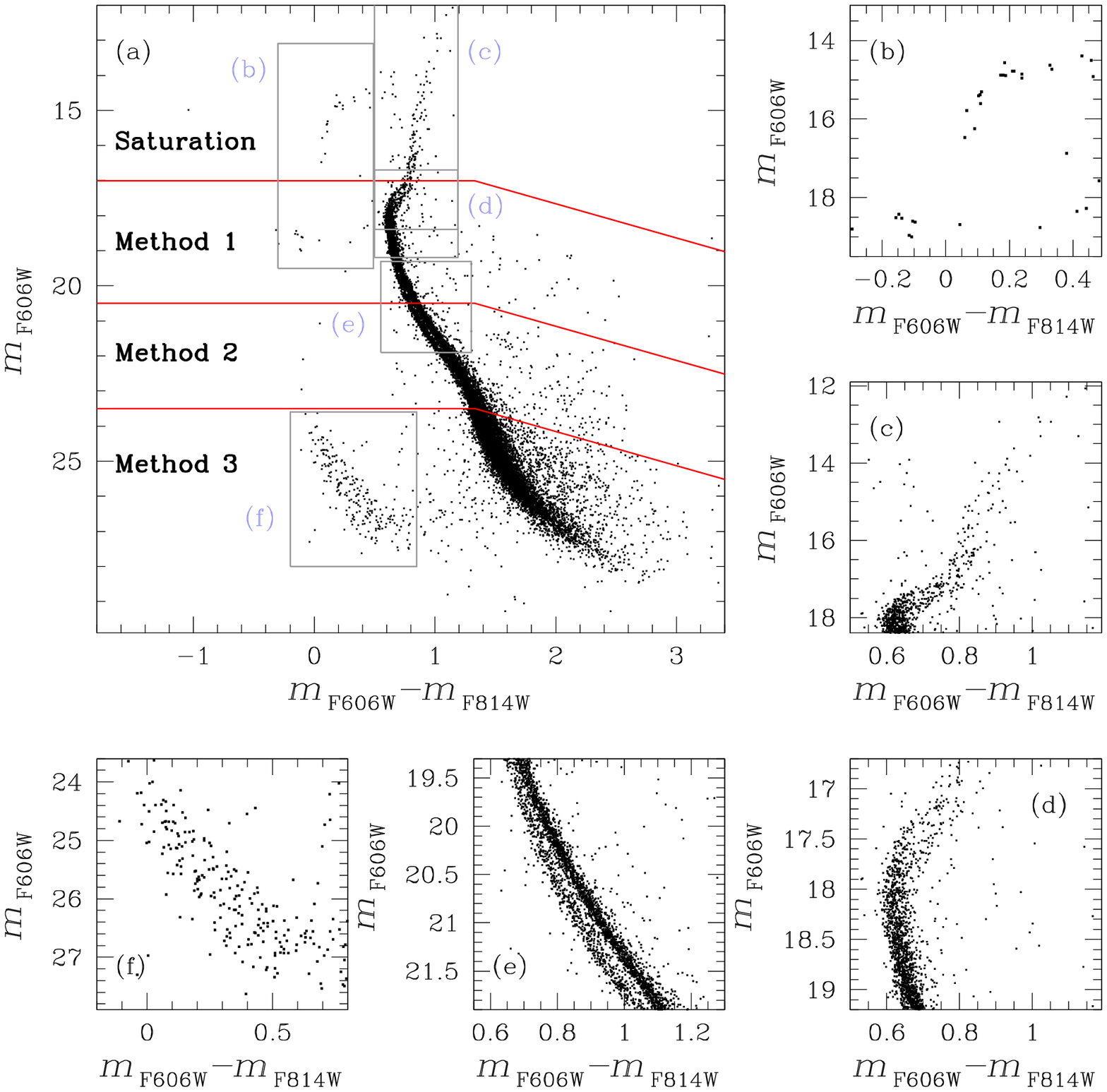}
  \caption{(a) Full $m_{\rm F606W}$ vs. $m_{\rm F606W}-m_{\rm F814W}$ CMD of $\omega$ Cen obtained by combining the best stars measured in the three photometric methods (see section 3.3). The transition between each photometric method is highlighted by red lines. All saturated stars are shown, with no selection. (b)-(f) Detail of the five regions that are outlined in (a). Moving clockwise, the panels show: (b) the HB, (c) the RGB, (d) the SGB, (e) the MS and (f) the WD cooling sequence.}
  \label{fig:CMD} 
 \end{figure*} 
 \end{centering}

Panel (a) of Figure \ref{fig:CMD} shows the full $m_{\rm F606W}$ vs. $m_{\rm F606W}-m_{\rm F814W}$ CMD obtained combining the best measured stars of the three different photometric methods, selected using the selection procedure described above. No selection cuts were applied to saturated stars. The three red lines define the regions within which stars are saturated in at least one filter, or for which photometry is obtained with one of the three methods. In panels (b)-(f), we show in detail the five regions that are outlined in panel (a) to provide an overview of specific evolutionary sequences. Clockwise from panel (b) to (f) we show the horizontal branch (HB), the red-giant branch (RGB), the sub-giant branch (SGB), the main sequence (MS) and the WD cooling sequences.

\section{Proper motions}
We computed the PMs using the technique developed by Bellini et al.\ (2014) and improved in Bellini et al.\ (2018) and Libralato et al.\ (2018). This is an iterative procedure that treats each image as a stand-alone epoch and can be summarised in two main steps: (1) transforms the stellar positions of each exposure into a common reference frame by means of a six-parameters linear transformation; (2) fit these transformed positions as a function of the epoch with a least-square straight line. The slope of this line, computed after several outlier-rejection stages, is a direct measurement of the PM.

Following Bellini et al.\ (2018), we excluded from the PM analysis the UVIS F275W\footnote{Filters bluer than F336W are affected by large colour-dependent positional residuals with respect to the UVIS distortion solution (Bellini et al.\ (2011)), and they are not suitable for high-precision astrometry.} and IR F110W and F160W exposures. This choice of excluding IR images from the PMs analysis is motivated by three reasons. First, our finding is done in filters F814W and F606W of UVIS (see Sect.\,3.3), which proved to have greatest signals for both WDs and low MS stars, and most importantly because UVIS has the highest angular resolution to avoid blends. Second, the higher resolving power and pixel size of UVIS, with respect to IR (39.75\,mas vs.\ 121\/mas), directly translates into an higher astrometric precision ($\sim$0.4\,mas, Bellini et al.\, 2011 for UVIS, vs. $\sim$1.2\,mas for IR, Anderson 2016).  Third, as IR and UVIS images essentially maps the very same epochs, IR would only have added noise to PMs measurements, mainly due to its lower resolving power, exposing to blends in these  relatively-high crowded fields.

We made use of stellar positions as measured by \texttt{KS2}'s method 1, which is best suited to high-precision PM analyses. As a common reference frame, we used star positions from the Gaia eDR3 catalogue, around 3 arcmin centered on the two fields F2 and F3. We transformed the \texttt{KS2} method-1 stellar positions, which are based on a reference frame that was obtained from the catalogue of a single WFC3/UVIS F814W exposure, by means of a six-parameter linear transformation. We defined an initial set of unsaturated reference stars, using the parameters described in section 3.6 to remove the poorly-measured stars, and we selected the likely cluster members on the basis of their positions in the CMD. The PM fitting and data rejection were performed exactly as described in Bellini et al.\ (2014), which provides a detailed description of the PM extraction and outlier rejection.

We iterated the procedure a few times in order to refine the reference-star list and the PM measurements. At the end of each iteration, we improved the reference-star list by removing all objects that have a large PM error or for which the PM is not consistent with the cluster's mean motion. The outlier-rejecting iterations stop when the number of reference stars differ by less than 2 from one iteration to the next. 

While for field F1 Bellini et al.\ (2018) made also use of archival data  collected several years earlier within \textit{HST} programs GO-9444 (PI: King, I. R.) and GO-10101 (PI: King, I. R.), here for fields F2 and F3, we have only the two epochs of GO14118 and GO14662, just $\sim$2\,years apart. Therefore our considerably shorter time-baseline ($\sim$2\,yr vs.\ $\sim$15\,yr)  directly translates into a proportionally inferior PM precision as compared with Bellini et al.\ (2018). 

The initial master list contained 42551 sources, 27885 ($\sim$65\%) of which had high-precision PMs. The missing 14666 sources were rejected at different iteration stages.  Our final catalogue is provided with the same set of quality and diagnostic parameters described in Bellini et al.\ (2014).

\begin{centering} 
 \begin{figure*}
  \includegraphics[width=15cm]{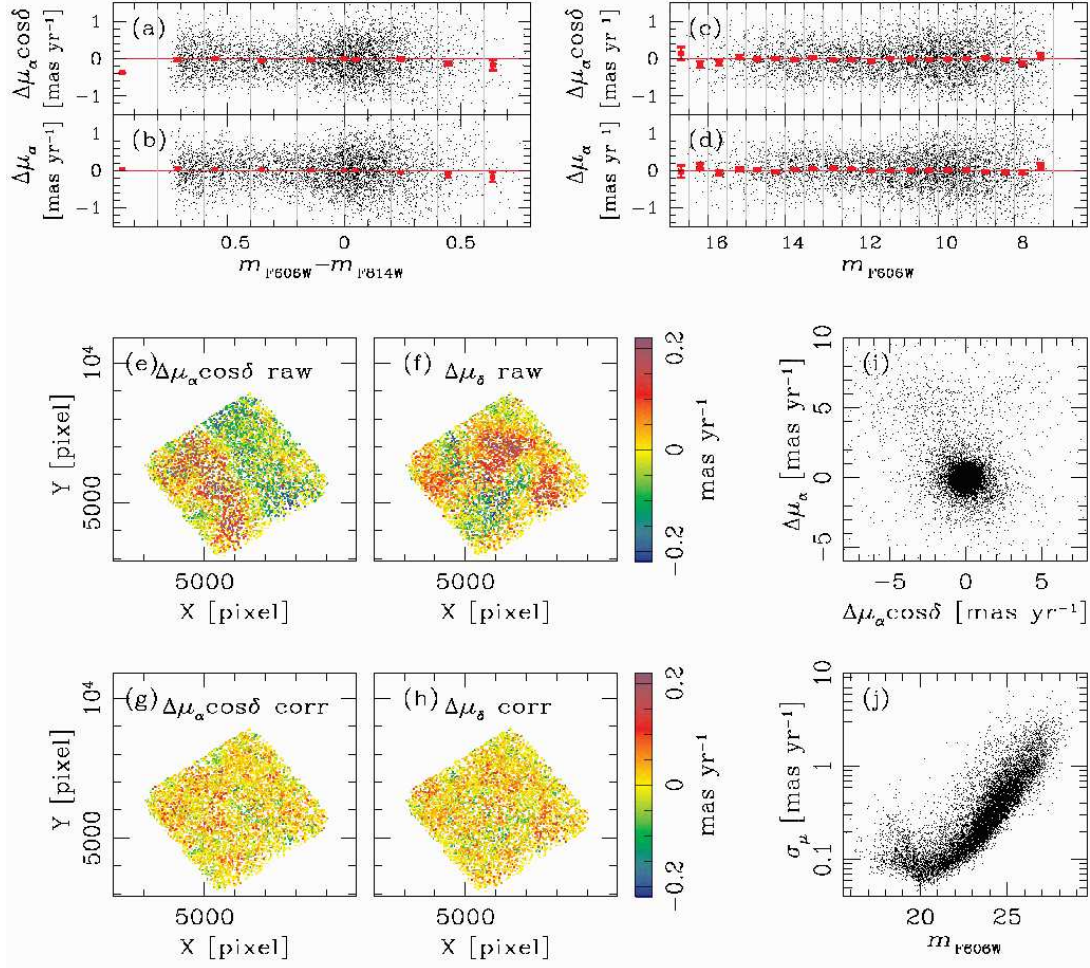}
  \caption{This figure illustrates the {\it a posteriori} procedure applied to the raw PM measurement for field F2. Panels (a) and (b) show that corrected PMs (in units of mas yr$^{-1}$) do not suffer from systematic effects as a function of stellar colour. Similarly, panels (c) and (d) show that corrected PMs do not suffer from systematic effects as a function of stellar magnitude. 
  In panels (e) and (f) we report the maps of the locally measured mean raw PM components of cluster members. Specifically, the deviation along $\mu_{\alpha} \cos\delta$ is in panel (e), and the deviation along $\mu_{\delta}$ is in panels (f). Each star is colour-coded according as shown by the vertical bar the immediate right of panel (f). Panels (g) and (h) show the maps of the locally-measured mean corrected PM of cluster members. We applied the same colour-scheme as in panels (e) and (f). Panel (i) shows the corrected proper motion diagram. Finally, panel (j) shows the corrected proper motion error in function of the F606W magnitude.}
  \label{fig:VPD2} 
 \end{figure*} 
 \end{centering}

\begin{centering} 
 \begin{figure*}
  \includegraphics[width=15cm]{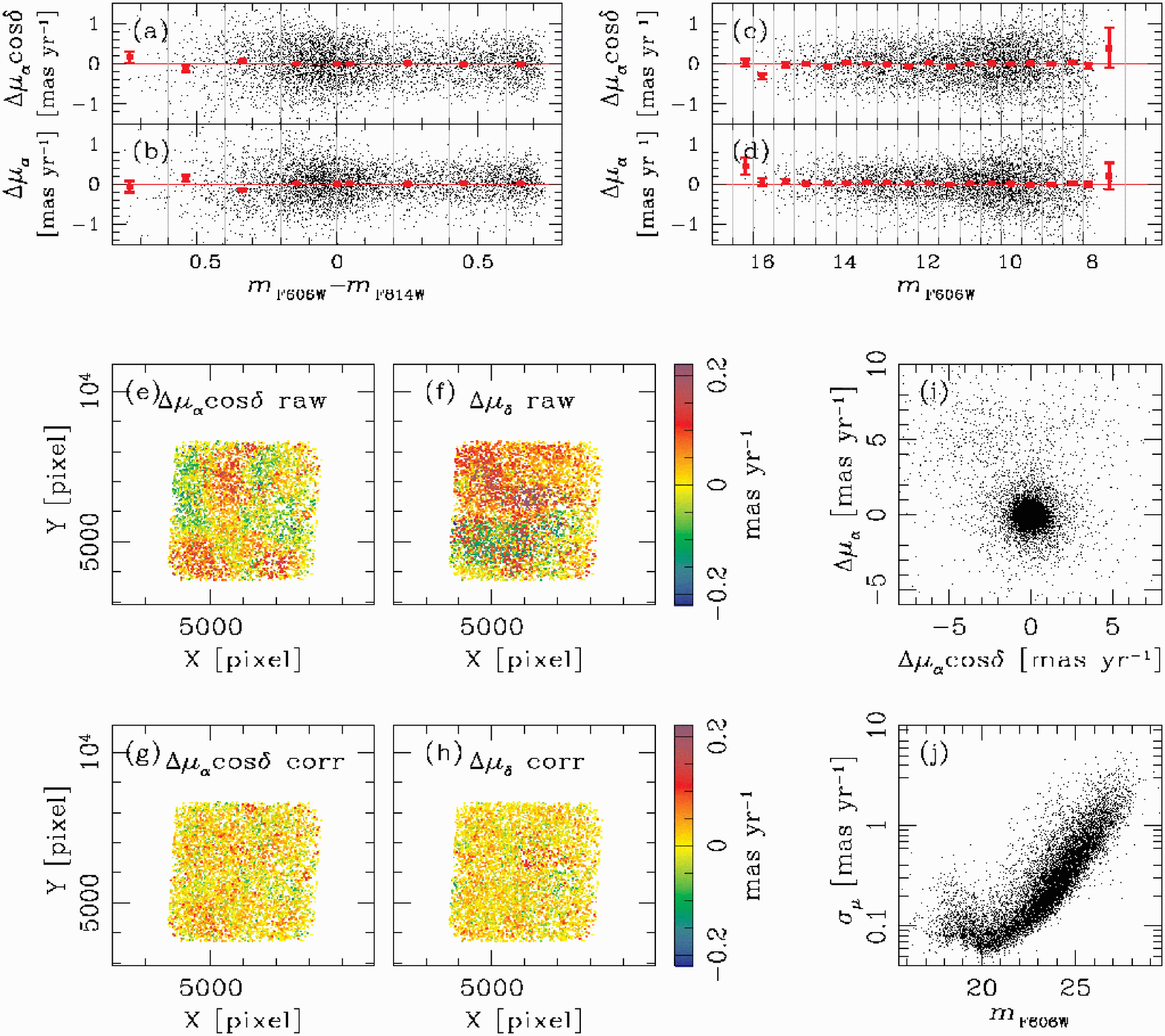}
  \caption{As in Figure \ref{fig:VPD2} but for field F3.}
  \label{fig:VPD3} 
 \end{figure*} 
 \end{centering}

Systematic errors in the PMs were corrected following the prescription of Bellini et al.\ (2014, Sec.\ 7.3 and 7.4) and Bellini et al.\ (2018). Figure \ref{fig:VPD2} and \ref{fig:VPD3} illustrate the correction procedure for field F2 and F3 respectively.

We started by selecting likely cluster members on the basis of their position on the PM diagram (within 1.5 mas yr$^{-1}$ from the bulk distribution) and rejecting all sources with a large PM error. Local PM corrections were applied as described in Section 7.4 of Bellini et al.\ (2014). In brief, systematic errors were mitigated "{\it a posteriori}", locally correcting the PM of each star according to the 2.5$\sigma$-clipped median value of the closest likely cluster members and within 0.5 $m_{\rm F606W}$ magnitudes from the target star (excluding the target star itself).

Panels (e) and (f) show the maps of the local median values obtained with the uncorrected (raw) components of the motion: $\Delta\mu_{\alpha} \cos\delta$ in panels (e) and $\Delta\mu_{\delta}$ in panels (f). Each point is a source, colour-coded according its locally-averaged PM value, as shown on the colour bar on the right-hand side of panels (e). Panels (g) and (h) show similar maps after the high-frequency variations are corrected. Points are colour-coded using the same colour scheme as panels (e) and (f).

We verified that after the correction, neither component of the corrected PM suffers from systematic effects due to stellar colour (panels (a) and (b)) and luminosity (panels (c) and (d)), dividing the stars into bins of fixed size in colour and magnitude, and evaluating the 3$\sigma$-clipped median value of the motion along $\Delta\mu_{\alpha} \cos\delta$ and $\Delta\mu_{\delta}$. The lack of systematic effect is clearly visible from panels (a), (b), (c) and (d), where the computed median values are shown as a red filled circles, with error bars. 

The quantities $\Delta\mu_{\alpha} \cos\delta$ and $\Delta\mu_{\delta}$ are in units of mas yr$^{-1}$ in all the panels. The associated errors of the mean are typically smaller than the size of the red circles. As a reference the red horizontal line shows the lack of systematic effects.

Panel (i) shows the PM diagram after the {\it a posteriori} correction. Since our reference list consists of cluster members, our PMs are relative to the cluster mean motion, and cluster members are represented by the bulk in the centre of the PM diagram. All other sources are foreground and background field sources.   Finally, panel (j) shows PM errors as a function of the $m_{\rm F606W}$ magnitude. 

\subsection{Membership probability}
To derive the membership probability, we followed a method based on PMs described by Balaguer-N\'u\~nez et al.\ (1998), Bellini et al.\ (2009) and Nardiello et al.\ (2018). The density function of cluster and field stars is modelled with an axisymmetric 2D Gaussian distribution centered respectively on the origin of the vector-point diagram (VPD; since PMs are computed relative to the cluster's bulk motion) and on the field proper motion centre. Cluster and field stars were selected on the basis of their position on the VPD. For each target star, the membership probability was estimated using a sub-sample of reference stars having a magnitude similar to those of the target. The frequency function for the $i$-th star of a cluster is:
\begin{center}
    $\Phi_c^{\nu} = \frac{\text{exp}\Big\{-\frac{1}{2}\Big[\frac{(\mu_{xi}-\mu_{xc})^2}{\sigma_{xc}^2+\epsilon_{xi}^2}+\frac{(\mu_{yi}-\mu_{yc})^2}{\sigma_{yc}^2+\epsilon_{yi}^2}\Big]\Big\}}{2\pi (\sigma_c^2+\epsilon_{xi}^2)^{1/2}(\sigma_c^2+\epsilon_{yi}^2)^{1/2}}$
\end{center}
where $(\mu_{xi},\mu_{yi})$ are the proper motion of the $i$-th stars, $(\mu_{xc},\mu_{yc})$ the cluster proper motion centre, ($\sigma_{xc},\sigma_{yc}$) the intrinsic proper motion dispersion of member stars, defined as the $68.27^{th}$ percentile of the $\mu_{xi}$ and $\mu_{yi}$ distribution, and $(\epsilon_{xi},\epsilon_{yi})$ the observed errors of the proper-motion component of the $i$-th star. 
Similar for the field:
\begin{center}
    $\Phi_f^{\nu} = \frac{\text{exp}\Big\{-\frac{1}{2(1-\gamma^2)}\Big[\frac{(\mu_{xi}-\mu_{xf})^2}{\sigma_{xf}^2+\epsilon_{xi}^2}-\frac{2\gamma(\mu_{xi}-\mu_{xf})(\mu_{yi}-\mu_{yf})}{(\sigma_{xf}^2+\epsilon_{xi}^2)^{1/2}(\sigma_{yf}^2+\epsilon_{yi}^2)^{1/2}}+\frac{(\mu_{yi}-\mu_{yf})^2}{\sigma_{xf}^2+\epsilon_{yi}^2}\Big]\Big\}}{2\pi (1-\gamma^2)^{1/2}(\sigma_{xf}^2+\epsilon_{xi}^2)^{1/2}(\sigma_{yf}^2+\epsilon_{yi}^2)^{1/2}}$
\end{center}
where $(\mu_{xf},\mu_{yf})$ is the field proper motion centre, ($\sigma_{xf},\sigma_{yf}$) the field intrinsic proper motion dispersion, defined as the $68.27^{th}$ percentile of the $\mu_{xi}$ and $\mu_{yi}$ distribution, and $\gamma$ the correlation coefficient:
\begin{center}
    $\gamma = \frac{(\mu_{xi}-\mu_{xf})(\mu_{yi}-\mu_{yf})}{\sigma_{xf}\sigma_{yf}}$
\end{center}
The distribution function of all the stars can be computed as follows:  
\begin{center}
    $\Phi = \Phi_c + \Phi_f = (n_c \cdot \Phi_c^{\nu}) + (n_f \cdot \Phi_f^{\nu})$
\end{center}
where $n_c$ and $n_f$ are the normalized number of stars for cluster and field ($n_c$ + $n_f$ = 1). Therefore, for the $i$-th star the resulting membership probability is
\begin{center}
    $P_c(i) = \frac{\Phi_c(i)}{\Phi(i)}$
\end{center}
Our evaluation of the membership probability does not consider the sources' spatial distribution, since our apertures are small enough to treat the member/field ratio as constant.

\begin{centering} 
 \begin{figure*}
  \includegraphics[width=15cm]{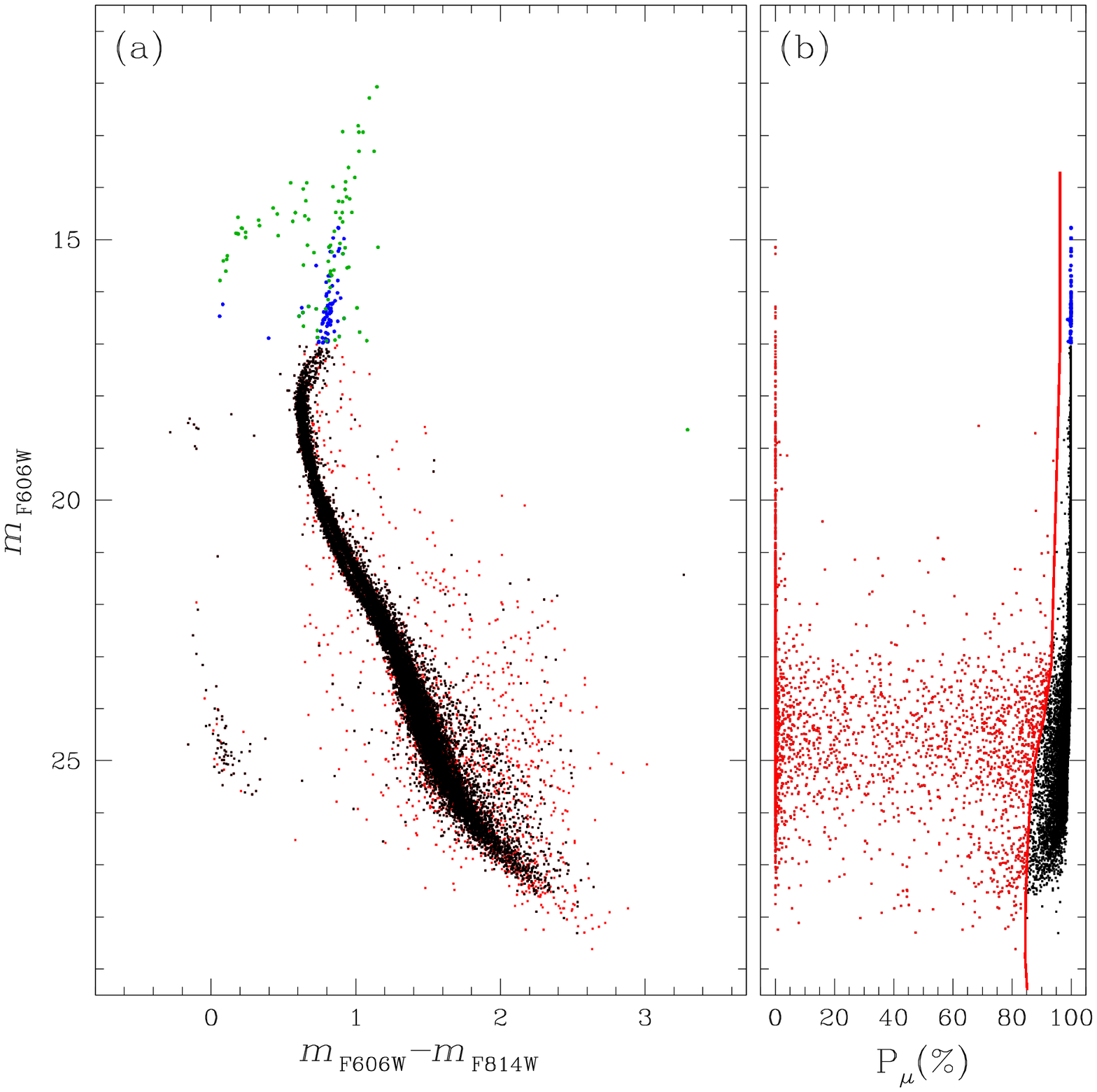}
  \caption{Probable-member selection. Only well-measured stars are shown. No quality selection cuts are applied to saturated stars. Panel (a) shows the $m_{\rm F606W}$ vs $m_{\rm F606W}-m_{\rm F814W}$ CMD. Saturated stars with no proper motion measurements and for which is impossible to estimate the membership probability are shown in green. Panel (b) presents the membership probability as a function of $m_{\rm F606W}$ magnitude, and the selection drawn by hand. In all panels, we highlight in red field stars and in black likely cluster member. Stars that are saturated in at least one filter are shown in blue.}
  \label{fig:Pmu} 
 \end{figure*} 
 \end{centering}

Figure \ref{fig:Pmu} shows an example of field-star decontamination based on membership probabilities. Poorly measured stars were removed using the parameters described in section 3.6, using a tighter selection than that in Figure \ref{fig:quality_sel}. No quality-selection cuts are applied for saturated stars. Panel (a) shows the $m_{\rm F606W}$ vs. $m_{\rm F606W}-m_{\rm F814W}$ CMD while panel (b) illustrates the membership distribution. The red line, drawn by hand, separates cluster members from field stars, which are represented in black and in red respectively in all the panels. We highlight in blue the stars that are saturated in at least one filter. Finally, saturated stars with no proper motion measurements and for which it is not possible to estimate the membership probability are represented in green. 

\section{The catalogue}
The catalogue consists of an astrometric- and several photometric files. Each file contains a description of the data and has the same number of lines, one for each source in the same order.

The astrometric file (ID\_XY\_RD.dat) contains the star ID, an identifier number associated to the field containing the star, stellar position both in X, Y (pixels) and R.A., DEC. (decimal degrees), followed by PM information and PM diagnostic as described in Bellini et al.\ (2014); the last 5 columns contain the proper motion and the associated  errors along R.A. and DEC. after the {\it a posteriori} correction, and the membership probability. Stars with no PM measurements, have a flag value of $-999.999$ for all PM related columns except for U$_{\rm ref}$ (a flag value that tells if a star was used as reference cluster member for the six-parameter linear transformation in the PMs evaluation), N$_{\rm found}$ and N$_{\rm used}$, which are instead flagged to $-999$.

For each filter, we provide a different file for each photometric method (e.g., F336W.m1.dat, F336W.m2.dat, or F336W.m3.dat for methods one, two and three, respectively) containing VEGAMAG magnitudes, quality parameters (RMS, QFIT, o, RADXS, N$_{\rm f}$ and N$_{\rm g}$) for each measured star. In addition, the method-one files also contain information about the local sky background, as well as a saturation flag to distinguish between unsaturated and saturated stars. For F606W and F814W filters, when a star is saturated or not found in the long exposures, its photometry is recovered from the short exposures. The photometry of saturated stars comes from the first-pass reduction.  

While for UVIS filters the saturation limit is fixed, to establish the saturation limit for IR filters, where final numbers are the results of multiple readings, it can be a hard task. For this reason, for IR filters we provide two different catalogues for each method, separating short and long exposures (e.g., F110W.m1.short.dat, F110W.m1.long.dat for short and long exposures respectively). 

If a star is not found in a given filter the VEGAMAG magnitude is flagged to $-99.999$ and the QFIT, o and RADXS parameters are flagged to 0. For stars measured in only one image, it is not possible to compute the RMS parameter, so its value is flagged to $-99.999$. Finally saturated stars have a flagged value of $-99.999$ for RMS and $0$ for the other parameters.  Tables 3-6 show an extract of the astrometric file and the three photometric files for the F814W filter.

\begin{centering} 
 \begin{figure*}
  \includegraphics[width=15cm]{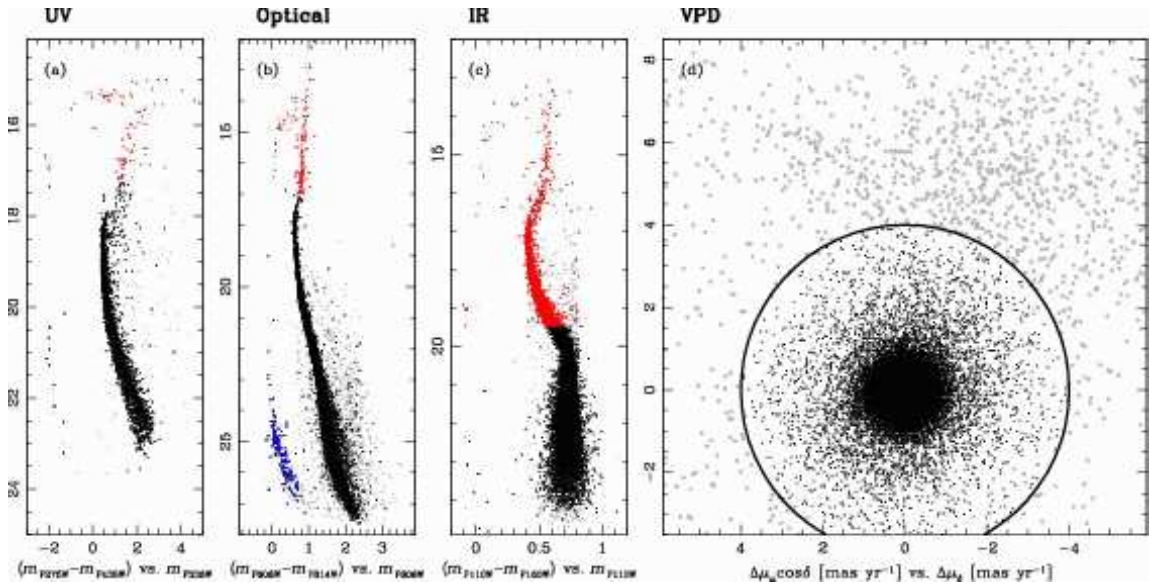}
  \caption{CMD for different interval of wavelength obtained after removing the poorly measured stars and correcting for differential reddening as in Bellini et al.\ (2017b). (a) $m_{\rm F336W}$ vs. $m_{\rm F275W}-m_{\rm F438W}$ ultraviolet CMD. (b) $m_{\rm F606W}$ vs. $m_{\rm F606W}-m_{\rm F814W}$ optical CMD. (c) $m_{\rm F110W}$ vs. $m_{\rm F110W}-m_{\rm F160W}$ near-infrared CMD. For panels (a), (b) and (c) saturated stars are in red while WDs are in blue and likely cluster members are represented in black while probable foreground and background field objects are in gray. Panel (d) shows the VPD of the relative PMs of the best measured stars in the two analysed fields, where we set a 4\,mas yr$^{-1}$ limit from the bulk distribution of the most probable cluster-members (black dots), and indicate background and foreground field sources with gray filled dots.
  }
  \label{fig:CMD2} 
 \end{figure*} 
 \end{centering}

\begin{table*}
\begin{centering}
\textbf{Table 3}\\
Extract of the Method-one F814W Photometry File\\
\smallskip
\end{centering}
\begin{tabular*}{1\textwidth}{@{\extracolsep{\fill}}c c c c c c c c c c}
\hline      
\hline
m$_{\rm F814W}$ & RMS & QFIT & o & RADXS & N$_{\rm f}$ & N$_{\rm g}$ & Sky (e$^-$) & rms Sky (e$^-$) & Sat flag\\
\hline
... & ... & ... & ... & ... & ... & ... & ... & ... & ...\\
 $19.9181$ &    $0.0101$ & $1.000$ &  $  0.00$ & $-0.0030$ &  $6$ &  $6$ & $ 522.7$ &  $289.4$ & $0$\\ 
 $19.0149$ &    $0.0047$ & $1.000$ &  $  0.00$ & $ 0.0001$ &  $8$ &  $8$ & $1102.3$ &  $634.0$ & $0$\\ 
 $21.1528$ &    $0.0390$ & $0.996$ &  $  0.00$ & $-0.0062$ &  $8$ &  $8$ & $ 245.4$ &  $138.7$ & $0$\\ 
 $20.8745$ &    $0.0072$ & $1.000$ &  $  0.00$ & $-0.0043$ &  $6$ &  $6$ & $ 252.3$ &  $131.0$ & $0$\\ 
 $20.2702$ &    $0.0080$ & $1.000$ &  $  0.00$ & $-0.0010$ &  $8$ &  $8$ & $ 397.2$ &  $206.7$ & $0$\\ 
 $22.3625$ &    $0.0286$ & $0.998$ &  $  0.00$ & $ 0.0081$ &  $7$ &  $7$ & $  91.2$ &  $ 31.4$ & $0$\\ 
 $21.4836$ &    $0.0095$ & $0.999$ &  $  0.00$ & $-0.0030$ &  $8$ &  $6$ & $ 161.1$ &  $ 71.0$ & $0$\\ 
 $22.8419$ &    $0.0127$ & $0.997$ &  $  0.00$ & $ 0.0092$ &  $7$ &  $6$ & $  77.5$ &  $ 22.7$ & $0$\\ 
 $22.2140$ &    $0.0201$ & $0.999$ &  $ 21.37$ & $ 0.0021$ &  $8$ &  $8$ & $ 104.8$ &  $ 35.2$ & $0$\\ 
 $22.5099$ &    $0.1260$ & $0.978$ &  $257.83$ & $ 0.0294$ &  $8$ &  $8$ & $ 110.4$ &  $ 46.9$ & $0$\\ 
 ... & ... & ... & ... & ... & ... & ... & ... & ... & ...\\
\hline
\end{tabular*}
\end{table*}
\begin{table}
\begin{centering}
\textbf{Table 4}\\
Extract of the Method-two F814W Photometry File\\
\smallskip
\end{centering}
\begin{tabular*}{.48\textwidth}{@{\extracolsep{\fill}}c c c c c c c}
\hline      
\hline
m$_{\rm F814W}$ & RMS & QFIT & o & RADXS & N$_{\rm f}$ & N$_{\rm g}$\\
\hline
... & ... & ... & ... & ... & ... & ...\\
 $19.9194$  &   $0.0097$ & $0.986$  & $  0.00$ & $ 0.0031$  & $6$  & $6$\\ 
 $19.0131$  &   $0.0056$ & $1.000$  & $  0.00$ & $-0.0009$  & $8$  & $8$\\ 
 $21.1541$  &   $0.0378$ & $0.995$  & $  0.00$ & $-0.0040$  & $8$  & $8$\\ 
 $20.8739$  &   $0.0074$ & $0.999$  & $  0.00$ & $-0.0044$  & $6$  & $6$\\ 
 $20.2685$  &   $0.0070$ & $1.000$  & $  0.00$ & $-0.0019$  & $8$  & $8$\\ 
 $22.3556$  &   $0.0351$ & $0.996$  & $  0.00$ & $ 0.0064$  & $7$  & $7$\\ 
 $21.4100$  &   $0.1351$ & $0.987$  & $  0.00$ & $ 0.0038$  & $8$  & $8$\\ 
 $22.8274$  &   $0.0241$ & $0.995$  & $  0.00$ & $ 0.0076$  & $7$  & $6$\\ 
 $22.1978$  &   $0.0226$ & $0.994$  & $ 14.96$ & $ 0.0031$  & $8$  & $8$\\ 
 $22.4234$  &   $0.1975$ & $0.971$  & $130.68$ & $ 0.0066$  & $8$  & $8$\\ 
 ... & ... & ... & ... & ... & ... & ...\\
\hline
\end{tabular*}
\end{table}
\begin{table}
\begin{centering}
\textbf{Table 5}\\
Extract of the Method-three F814W Photometry File\\
\smallskip
\end{centering}
\begin{tabular*}{.48\textwidth}{@{\extracolsep{\fill}}c c c c c c c}
\hline      
\hline
m$_{\rm F814W}$ & RMS & QFIT & o & RADXS & N$_{\rm f}$ & N$_{\rm g}$\\
\hline
... & ... & ... & ... & ... & ... & ...\\
 $19.9477$  &    $0.0189$ & $0.993$    & $ 0.00$ & $ 0.0154$  & $6$  & $6$\\ 
 $19.0156$  &    $0.0122$ & $1.000$    & $ 0.00$ & $ 0.0022$  & $8$  & $8$\\ 
 $21.1912$  &    $0.0092$ & $1.000$    & $ 0.00$ & $ 0.0100$  & $8$  & $8$\\ 
 $20.8616$  &    $0.0109$ & $1.000$    & $ 0.00$ & $ 0.0182$  & $8$  & $8$\\ 
 $20.2678$  &    $0.0077$ & $1.000$    & $ 0.00$ & $-0.0008$  & $8$  & $8$\\ 
 $22.4082$  &    $0.0372$ & $0.999$    & $ 0.00$ & $ 0.0260$  & $8$  & $8$\\ 
 $21.4870$  &    $0.0502$ & $0.999$    & $ 0.00$ & $ 0.0231$  & $8$  & $7$\\ 
 $22.8921$  &    $0.0433$ & $1.000$    & $ 0.00$ & $ 0.0572$  & $8$  & $8$\\ 
 $22.2390$  &    $0.0410$ & $0.998$    & $ 3.32$ & $ 0.0162$  & $8$  & $8$\\ 
 $22.6223$  &    $0.0556$ & $0.999$    & $20.22$ & $ 0.0608$  & $8$  & $8$\\ 
 ... & ... & ... & ... & ... & ... & ...\\
\hline
\end{tabular*}
\end{table}
A visual summary of the catalogue is given in Figure \ref{fig:CMD2} for three different CMDs, obtained using filters that span different intervals of wavelength: $m_{\rm F336W}$ vs. $m_{\rm F275W}-m_{\rm F438W}$ (ultraviolet filters) in panel (a), $m_{\rm F606W}$ vs. $m_{\rm F606W}-m_{\rm F814W}$ (optical filters) in panel (b) and $m_{\rm F110W}$ vs. $m_{\rm F110W}-m_{\rm F160W}$ (near-infrared filters) in panel (c). Poorly measured stars are removed using the photometric parameters described in section 3.6 and following the selection illustrated in Figure \ref{fig:quality_sel}. Probable cluster members and background sources are separated using the membership probability, and are represented respectively with black and gray dots in panels (a), (b) and (c) of Figure \ref{fig:CMD2}. We corrected the photometry for differential reddening following the method described in Bellini et al.\ (2017b, Sec.\ 3), which is an evolution of procedures described in Sarajedini et al.\ (2007). Saturated stars are represented in red while WDs are in blue. Panel (d) shows the VPD of the relative PMs, after the {\it a posteriori} correction, of the two analysed fields. Likely cluster-member, within 4 mas yr$^{-1}$ from the bulk distribution, are represented with black dots, while background and foreground field sources are represented with gray filled dots. 

Together with the astro-photmetric catalogue, we also release for each of the two fields the atlases, i.e., a view of the field through stacked images. We produce those in two versions: sampled at 1$\times$- and at 2$\times$-supersampled pixels. These stacked images are in standard \texttt{fits} format and contain in their headers the astrometric WCS solution linked to Gaia\,eDR3.  For each field, we provide one single stack image each for filters F275W, F336W and F438W, and two stack images for each of F606W, F814W, F110W and F160W, separating short- and long-exposure images.

\begin{centering} 
 \begin{figure*}
  \includegraphics[width=5.85cm]{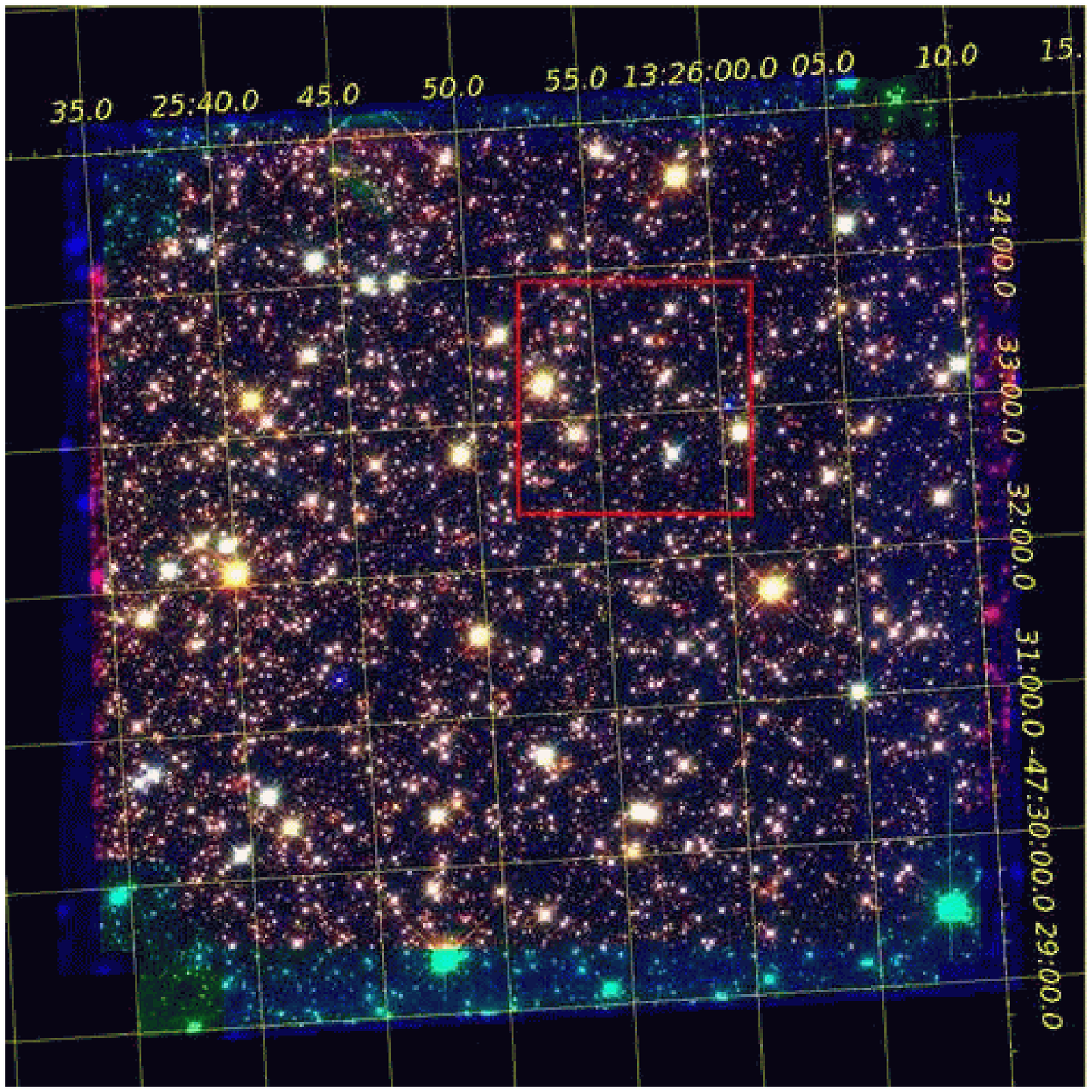}
  \includegraphics[width=5.85cm]{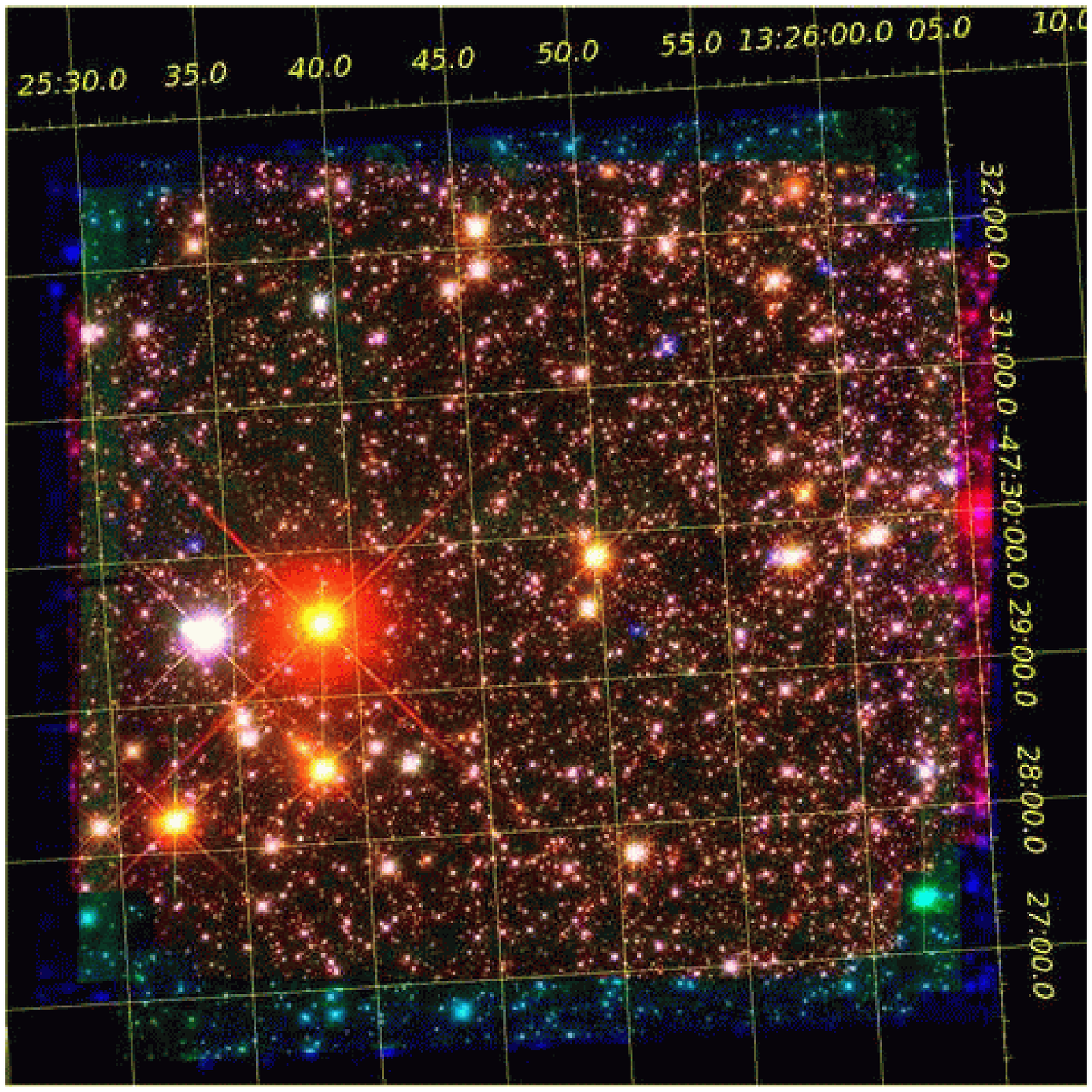}
  \includegraphics[width=5.85cm]{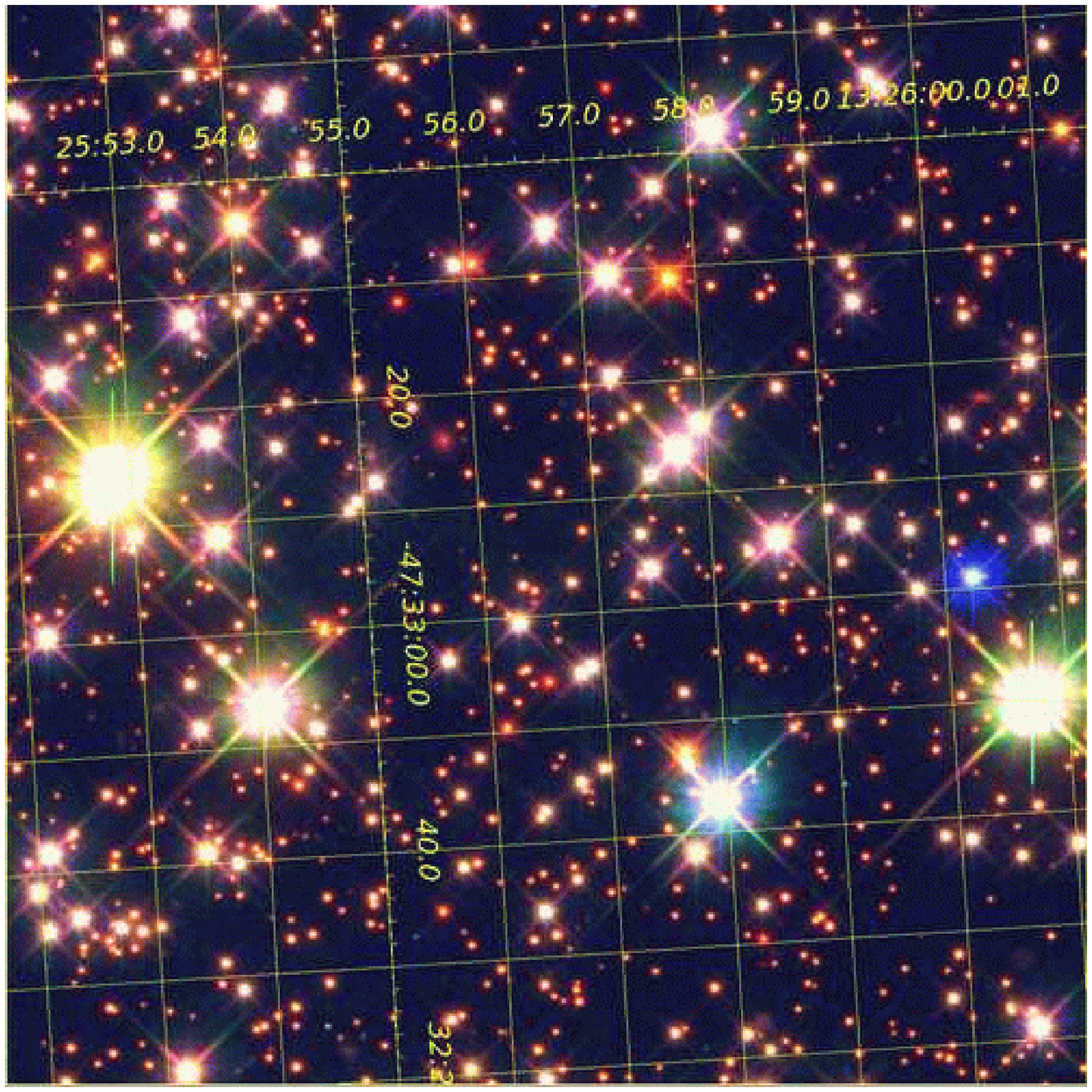}
  \caption{Three-colour images for field F2 (left) and F3 (middle) with an ICRS grid over-imposed for reference (in yellow). 
  In the right panel we show the zoom-in indicated with the red box in field F2. The red, green, and blue channels are filled by F110W, F606W and F336W, respectively.}
  \label{fig:ATLASes} 
 \end{figure*} 
 \end{centering}

To give a visual sense of the stacks, we show in Figure \ref{fig:ATLASes} three-colour images for field F2 (left) and F3 (middle). An ICRS grid is overimposed in each images for reference. In the right panel we show a zoomed region, at a scale that shows the individual pixels, of the field F3.

\section{Summary and Conclusions}

As part of the delivery of our large programme on $\omega$ Cen
we are committed to release astro-photometric catalogues of all our studied fields.  Along with this publication, we present and publicly release \textit{(i)} the astro-photometric catalogue, and \textit{(ii)} the multi-band atlases for the remaining two (out of three) WFC3 parallel fields F2 and F3 (indicated in Figure \ref{fig:fields}), which were not previously released (field F1 is the only one that has been released so far, Bellini et al.\ 2018). 

The catalogues provide stellar positions, PMs and 
PM diagnostic quantities, magnitudes and quality parameters.  Each file contains a header with a short description of the data it contains.
Together with the catalogue, we release atlases in each filter. These are stacked images available in two versions: one in original pixel size, and a version super-sampled by a factor 2. Both versions have headers containing the astrometric solutions with keywords in the WCS. We make images publicly available at our 
\texttt{url}\footnote{https://web.oapd.inaf.it/bedin/files/PAPERs$\_$eMATERIALs/\-wCen\_HST\_LargeProgram/P04/}
and as supplementary electronic on-line material of this journal. 
Upon reasonable request, we could also provide light curves for each filter of individual sources.  

The scientific exploitation of the present catalogue 
has great potential. The most immediate and  simple application would be to select interesting proper-motion members in any of the identified mPOPs sequences or in the binary sequence for detailed spectroscopic follow-up investigation. The catalogue is an \textit{HST} legacy and can provide an early epoch for future and astrometric campaigns, which can extend the time-baseline and therefore enable more accurate differential internal kinematic investigations among the different mPOPs of $\omega$\,Cen, as well as many other unforeseeable uses.  

In future publications, we will use also F1, the central field (Bellini et al.\ 2017a,b,c), and two other fields from another on-going program 
(GO-16247, PI: Scalco), to investigate the spatial properties of $\omega$\,Cen, in particular 
searching for radial gradients in: 
\textit{(i)  } the multiple populations (following analysis for F1 in Paper\,I); 
\textit{(ii) } the internal differential kinematic such as anisotropy and deviation from energy equipartion (following prescription of Paper\,II), and 
\textit{(iii)} the global kinematic properties, searching for possible systemic motions in the plane of the sky for the different sub-populations 
and as function of their stellar components at the various masses (following methodology of Paper\,III).

\section*{Acknowledgements}
Michele Scalco and Luigi Rolly Bedin acknowledge support by MIUR under PRIN program \#2017Z2HSMF. Domenico Nardiello acknowledges the support from the French Centre National d'Etudes Spatiales (CNES). Andrea Bellini acknowledges support from HST grants GO-14118+GO-14662.

\section*{Data Availability}

The data underlying this article were accessed from the Mikulski Archive for Space Telescopes (MAST), available at https://archive.stsci.edu/hst/search.php. All data come from \textit{HST} program GO-14118+GO-14662 (P.I. Bedin). The full list of observations are reported in Table 1.


\bibliographystyle{mnras}


\begin{landscape}
\begin{centering}
\textbf{Table 6}\\
Extract of the Astrometric File\\
\smallskip
\end{centering}
\centering
\scriptsize
\begin{tabular}{c c c c c c c c c c c c c c c}
\hline      
\hline
ID & Field id. & X & Y & R.A. & Decl. & $\Delta\mu_{\alpha}^{r} \cos\delta$ & $\Delta\mu_{\delta}^{r}$ & $\sigma_{\mu_{\alpha}^{r} \cos\delta}$ & $\sigma_{\mu_{\delta}^{r}}$ & $\sigma_x$ & $\sigma_y$ & $\chi_x^2$ & $\chi_y^2$ & $\rightarrow$\\
\hline
   ... & ... & ... & ... & ... & ... & ... & ... & ... & ... & ... & ... & ... & ... & ...\\
$31864$ & $3$ & $3938.8330$ & $2849.8887$ & $201.39965820312$ & $-47.46526718140$    & $0.44144$   & $-0.20060$    & $0.06848$    & $0.13760$    & $0.00186$    & $0.00373$    & $0.8387$    & $3.3881$ & .. \\
$31865$ & $3$ & $3922.3325$ & $2792.9084$ & $201.39933776855$ & $-47.46464920044$    & $0.46440$   & $ 0.12772$    & $0.13208$    & $0.15728$    & $0.00367$    & $0.00437$    & $1.2182$    & $1.7280$ & .. \\
$31866$ & $3$ & $3856.4814$ & $2822.9153$ & $201.39828491211$ & $-47.46502685547$    & $0.14316$   & $ 0.17064$    & $0.11548$    & $0.15248$    & $0.00307$    & $0.00406$    & $1.0918$    & $1.9037$ & .. \\
$31867$ & $3$ & $3857.5540$ & $2852.8218$ & $201.39834594727$ & $-47.46535491943$    & $0.03576$   & $-0.06868$    & $0.10500$    & $0.10940$    & $0.00285$    & $0.00297$    & $1.2943$    & $1.4046$ & .. \\
$31868$ & $3$ & $3865.2600$ & $2800.5972$ & $201.39840698242$ & $-47.46477508545$    & $0.65080$   & $ 0.52796$    & $0.16276$    & $0.25592$    & $0.00434$    & $0.00683$    & $0.8630$    & $2.1339$ & .. \\
$31869$ & $3$ & $3885.8447$ & $2808.9851$ & $201.39874267578$ & $-47.46485137939$    & $0.08020$   & $-0.22192$    & $0.13824$    & $0.24332$    & $0.00375$    & $0.00660$    & $1.0195$    & $3.1589$ & .. \\
$31870$ & $3$ & $3887.8281$ & $2859.1301$ & $201.39883422852$ & $-47.46540451050$    & $0.80596$   & $ 0.08448$    & $0.29240$    & $0.59520$    & $0.00893$    & $0.01819$    & $1.2478$    & $5.1708$ & .. \\
$31871$ & $3$ & $3897.1392$ & $2843.9988$ & $201.39897155762$ & $-47.46522903442$    & $0.02488$   & $-0.08076$    & $0.16104$    & $0.29644$    & $0.00440$    & $0.00810$    & $1.0012$    & $3.3923$ & .. \\
$31872$ & $3$ & $3899.6646$ & $2882.8450$ & $201.39904785156$ & $-47.46565628052$    & $0.47840$   & $-1.95504$    & $0.26004$    & $0.33260$    & $0.00724$    & $0.00926$    & $1.6503$    & $2.7002$ & .. \\
$31873$ & $3$ & $3903.2605$ & $2879.6541$ & $201.39912414551$ & $-47.46562194824$    & $0.80652$   & $ 0.49452$    & $0.35804$    & $0.53260$    & $0.00989$    & $0.01471$    & $1.8483$    & $4.0904$ & .. \\
   ... & ... & ... & ... & ... & ... & ... & ... & ... & ... & ... & ... & ... & ... & ...\\
\hline
$\rightarrow$ & U$_{\rm ref}$ & N$_{\rm found}$ & N$_{\rm used}$ & $\Delta$Time & err$_{\mu_{\alpha}^{r} \cos\delta}$ & err$_{\mu_{\delta}^{r}}$ & err$_x$ & err$_y$ & $\Delta\mu_{\alpha}^{c} \cos\delta$ & $\Delta\mu_{\delta}^{c}$ & $\sigma_{\mu_{\alpha}^{c} \cos\delta}$ & $\sigma_{\mu_{\delta}^{c}}$ & 
$P_\mu$ \\
\hline
... & ... & ... & ... & ... & ... & ... & ... & ... & ... & ... & ... & ... & ...\\
... &    $1$ &   $34$ &   $34$ &    $2.00439$ &    $0.05672$ &    $0.11996$ &    $0.00344$  &   $0.00309$ &   $ 0.43699$ &   $-0.24423$ &   $ 0.07686$ &    $0.13885$ &  $99.9909$\\
... &    $1$ &   $29$ &   $28$ &    $2.00439$ &    $0.14156$ &    $0.12628$ &    $0.00393$  &   $0.00353$ &   $ 0.44001$ &   $ 0.02924$ &   $ 0.15118$ &    $0.16260$ &  $99.9338$\\
... &    $1$ &   $31$ &   $31$ &    $2.00439$ &    $0.08708$ &    $0.10688$ &    $0.00381$  &   $0.00233$ &   $ 0.12795$ &   $ 0.09779$ &   $ 0.10001$ &    $0.13951$ &  $99.9465$\\
... &    $1$ &   $37$ &   $36$ &    $2.00439$ &    $0.07096$ &    $0.07740$ &    $0.00273$  &   $0.00193$ &   $ 0.03030$ &   $-0.20915$ &   $ 0.08265$ &    $0.15485$ &  $99.9776$\\
... &    $0$ &   $23$ &   $22$ &    $2.00439$ &    $0.17916$ &    $0.19156$ &    $0.00640$  &   $0.00598$ &   $ 0.57807$ &   $ 0.46604$ &   $ 0.20189$ &    $0.20722$ &  $99.0765$\\
... &    $1$ &   $25$ &   $22$ &    $2.00439$ &    $0.11404$ &    $0.18928$ &    $0.00608$  &   $0.00500$ &   $ 0.00110$ &   $-0.34661$ &   $ 0.14487$ &    $0.22597$ &  $99.8914$\\
... &    $0$ &   $20$ &   $20$ &    $2.00439$ &    $0.33244$ &    $0.94336$ &    $0.01488$  &   $0.03188$ &   $ 0.78886$ &   $ 0.02099$ &   $ 0.33755$ &    $0.94631$ &  $97.8341$\\
... &    $1$ &   $24$ &   $24$ &    $2.00439$ &    $0.11276$ &    $0.28732$ &    $0.00741$  &   $0.00650$ &   $-0.04496$ &   $-0.10662$ &   $ 0.14182$ &    $0.29259$ &  $99.6259$\\
... &    $0$ &   $23$ &   $22$ &    $2.00439$ &    $0.29160$ &    $0.38592$ &    $0.00831$  &   $0.01076$ &   $ 0.46639$ &   $-1.97989$ &   $ 0.29785$ &    $0.39022$ &  $72.5362$\\
... &    $0$ &   $23$ &   $23$ &    $2.00439$ &    $0.31708$ &    $0.65620$ &    $0.01331$  &   $0.02024$ &   $ 0.77549$ &   $ 0.52001$ &   $ 0.32391$ &    $0.65863$ &  $97.3289$\\
... & ... & ... & ... & ... & ... & ... & ... & ... & ... & ... & ... & ... & ...\\  
\hline
\end{tabular}
\end{landscape}

\bsp	
\label{lastpage}
\end{document}